\documentclass[12pt,a4paper]{article}

\usepackage{amsmath,amsthm,amssymb}
\usepackage{graphicx}
\date{}
\usepackage{url}
\urldef{\mailsa}\path|{alfred.hofmann, ursula.barth, ingrid.haas, frank.holzwarth,|
\urldef{\mailsb}\path|anna.kramer, leonie.kunz, christine.reiss, nicole.sator,|
\urldef{\mailsc}\path|erika.siebert-cole, peter.strasser, lncs}@springer.com|

\begin{document}

\title{A new construction of differentially 4-uniform permutations over $F_{2^{2k}}$}

\author{Jie Peng\thanks{J. Peng is with the School of Mathematics and Statistics, Central China Normal University, Luoyu Road \#152, Wuhan
430079, P. R. China (e-mail:~jiepeng@mail.ccnu.edu.cn). Now he is visiting Temasek Laboratories, National University of Singapore.}\and
Chik~How~Tan and Qichun~Wang\thanks{C. H. Tan and Q. C. Wang are with Temasek Laboratories, National University of Singapore, 5A Engineering Drive 1, 09-02, 117411 Singapore. (e-mails:~\{tsltch,tslwq\}@nus.edu.sg).} }

%
\maketitle
\begin{abstract}
 Permutations over $F_{2^{2k}}$ with low differential uniform, high algebraic degree and high nonlinearity are of great cryptographical importance since they can be chosen as the substitution boxes (S-boxes) for many block ciphers. A well known example is that the Advanced Encryption Standard (AES) chooses a differentially 4-uniform permutation, the multiplicative inverse function, as its S-box. In this paper, we present a new construction of differentially 4-uniformity permutations over even characteristic finite fields and obtain many new CCZ-inequivalent functions. All the functions are switching neighbors in the narrow sense of the multiplicative inverse function and have the optimal algebraic degree and high nonlinearity.

\end{abstract}



\section{Introduction}

An S-box with $n$ input bits and $m$ output bits is an $(n,m)$-function $f: F_{2^n}\rightarrow F_{2^m}$. For ease of implementation, they are usually chosen to be permutations over $F_{2^n}$ for even integer $n$. The concept of differential uniformity is used to measure the ability of the function to resist the so called differential attack on the design of block encryption algorithm, which was presented by Biham and Shamir \cite{Biham}. The value of differential uniformity of S-boxes should be as low as possible. It is well known that the minimal value of differential uniformity equals 2 for an $(n,n)$-function, and if this value is achieved, then the function is called almost perfect nonlinear (APN). Several classes of APN functions are found when $n$ is odd (see \cite{Bracken0,Edel} and the references therein). But when $n$ is even, only one sporadic example $x^3+Tr(x^9)$ of APN permutation for $n=6$ has been found in reference \cite{Dillon1} and it is still an open problem whether there exist APN permutations for even $n\geq 8$. However, S-boxes are often required to be permutations over finite fields with even characteristic for efficient software implementation. Therefore, differentially 4-uniform permutations become ideal candidates for S-boxes. For example, AES chooses the multiplicative inverse function $x^{2^n-2}$ as its S-box, which is differentially 4-uniform with optimal algebraic degree and known maximum nonlinearity. Up to now, several classes of differentially 4-uniform permutations over finite fields with even characteristic have been presented, which are listed below for the readers' convenience.

\begin{itemize}
\item  [1)] Gold function \cite{Gold}: $x^{2^i+1}$, where $n=2k$, $k$ is odd and $\gcd(n, i)=2$;

\item [2)] Kasami function \cite{Kasami}: $x^{2^{2i}-2^i+1}$, where $n=2k$, $k$ is odd and $\gcd(n, i)=2$;

\item [3)] Multiplicative inverse function \cite{Nyberg}: $x^{-1}$ (as usual $0^{-1}:=0$), where $n$ is even;

\item [4)] Bracken-Leander function \cite{Bracken1}: $x^{2^{2k}+2^k+1}$, where $n=4k$ and $k$ is odd;

\item [5)] Binomial function \cite{Bracken2}: $\alpha x^{2^s+1}+\alpha^{2^k}x^{2^{-k}+2^{k+s}}$, where $n=3k$, $k$ even, $k/2$ odd, $\gcd(n, s)=2$, $3| k+s$ and $\alpha$ is a primitive element of $F_{2^n}$;

\item [6)] Qu-Tan-Tan-Li function \cite{Qu}: The 1st class: $x^{-1}+Tr(x+(x+1)^{-1})$; The 2nd class: $x^{-1}+Tr(x^{-d}+(x^{-1}+1)^d)$, where $d=3(2^t+1)$ and $2\leq t\leq n/2-1$;

\item [7)] Zha-Hu-Sun function \cite{Zha}: The 1st class: $x^{-1}+t(x^{2^s}+x)^{2^{n}-1}+t$, where $t\in F_{2^s}$, $s|n$, and $s$ is even, or $s=1, 3$ and $n/2$ is odd; The 2nd class: $t_1x^{-1}+(t_1+1)x^{-1}(x^{2^s}+x)^{2^{n}-1}+t_2(x^{2^s}+x)^{2^{n}-1}+t_2$, where $t_1,t_2\in F_{2^s}$ with $Tr(t_1^{-1})=1$, $s|n$, $s$ is even and $n/s$ is odd.

\item [8)] Tang-Carlet-Tang function \cite{Tang}: $(x+\delta_T(x))^{-1}$, where $\delta_T$ is the indicator function of the set $T\subseteq F_{2^{2k}}$ with $k\geq 3$, which satisfies

(1) if $x\in T$, then $x+1\in T$, and

(2) if $x\in T$, then $Tr\bigg(\frac{1}{x}\bigg)=Tr\bigg(\frac{1}{x+1}\bigg)=1$.

\end{itemize}

Furthermore, functions used as S-boxes should have other good cryptographic properties as well in order to resist other types of attacks. For instance, to resist the higher order differential attack \cite{Knudsen} and linear attack \cite{Matsui}, the algebraic degree and nonlinearity of S-boxes should be as high as possible. It is known that for an $(n, n)$ permutation polynomial, the highest possible degree is $n-1$. However, as is mentioned in some papers (such as \cite{Qu,Tang}), though the first 5 classes of functions all have the known highest nonlinearity, except the Kasami and multiplicative inverse functions, the algebraic degrees of the other 3 classes are only 2 or 3, which are too low to be used as S-boxes. Moreover, only the multiplicative inverse function exists for any even integer $n$. Though the multiplicative inverse function has optimal algebraic degree, and has been used as the S-box of the AES, it may also risk a threat for algebraic attacks \cite{Tang}. Therefore, it is urgent to find more classes of differentially 4-uniform permutations, which have high nonlinearity and algebraic degree simultaneously.

A powerful secondary construction method called switching method was presented to obtain new APN functions by changing some of the coordinate functions of a given function. For instance, the famous APN function $x^3+Tr(x^9)$ was obtained via changing one of the coordinate function of the Kasami function. Besides, this method can be used to construct cryptographic significant differentially 4-uniform permutations as well (see \cite{Qu,Tang,Tan,Zha} and the references therein). In \cite{Zha}, the authors succeeded to construct differentially 4-uniform permutations via modifying the multiplicative inverse function on some subfield of $F_{2^n}$. In \cite{Qu}, it was proved that given a so called preferred function $R(x)$, then $x^{-1}+Tr(R(x^{-1})+R(x^{-1}+1))$ is a differentially 4-uniform permutation. This construction was further investigated in \cite{Qu1}. In \cite{Tang}, the authors presented a new construction by permuting the multiplicative inverse function, which could produce many CCZ-inequivalent differentially 4-uniform permutations.
The functions in these references were proved to have optimal algebraic degree and high nonlinearity. The purpose of this paper is to construct more classes of differentially 4-uniform permutations with high nonlinearity and algebraic degree for any even integer $n\geq 6$, which can provide more choices for S-boxes.

In \cite{Qu}, the authors obtained the roots $x_1, x_2$ of a quadratic equation $x^2+\frac{1}{b}x+\frac{1}{b(b+1)}=0$ given that it did have a solution, and used some properties of $x_1, x_2$ to help construct differentially 4-uniform functions. In this paper, we study and get some relationships on the values $Tr(x)$ and $Tr\bigg(\frac{1}{x+1}\bigg)$ of these two roots. Based on that, we modify the multiplicative inverse function on some subsets of $F_{2^n}$ and obtain a new construction of differentially 4-uniform permutations. Moreover, all the newly discovered functions have optimal algebraic degree and high nonlinearity.


The rest of this paper is organized as follows. In the next section we give some basic knowledge and results. In section 3 we present our new construction of differentially 4-uniform permutations. And we analyze the other cryptography of the new functions in Section 4. Finally, Section 5 concludes the paper.

\section{Preliminaries}

An $(n,n)$-function $F: F_{2^n}\rightarrow F_{2^n}$ can be uniquely represented as a polynomial
\begin{eqnarray*} F(x)=\sum_{i=0}^{2^n-1}a_ix^i, a_i\in F_{2^n}.
\end{eqnarray*} We denote the algebraic degree of $F$ by $\deg(F)$, which is defined to be the maximum 2-weight of $i$ (i.e., the number of 1's in the 2-adic representation of $i$) such that $a_i\neq 0$. $F$ is called an affine function if $\deg(F)\leq 1$. It is well known that $\deg(F)\leq n-1$ when $F$ is a permutation over $F_{2^n}$. And if the equality holds, we say $F$ has optimal algebraic degree.

Let $Tr_k^n(x)$ be the trace map from $F_{2^n}$ to its subfield $F_{2^k}$, i.e., $Tr_k^n(x)=\sum_{i=0}^{n/k-1}x^{2^{ik}}$. And the absolute trace function ($k=1$) is denoted by $Tr(x)$ for simplicity.

For an $(n,n)$-function $F$, its Walsh transform is defined by
\begin{eqnarray*} F^{\mathcal{W}}(a,b):=\sum_{x\in F_{2^n}}(-1)^{Tr(ax+bF(x))}, a\in F_{2^n}, b\in F_{2^n}^*.
\end{eqnarray*} And the multisets $\{*~F^{\mathcal{W}}(a,b)~|~(a, b)\in F_{2^n}\times F_{2^n}^*~*\}$ and $\{*~|F^{\mathcal{W}}(a,b)|~|~ (a, b)\in F_{2^n}\times F_{2^n}^*~*\}$ are called Walsh spectrum and extended Walsh spectrum of $F$. The nonlinearity of $F$, denoted by $NL(F)$, is related to its extended Walsh spectrum:
\begin{eqnarray*}NL(F)=2^{n-1}-\frac{1}{2}\max_{(a, b)\in F_{2^n}\times F_{2^n}^*}|F^{\mathcal{W}}(a,b)|.
\end{eqnarray*} When $n$ is odd, it has been proved that $NL(F)\leq 2^{n-1}-2^{\frac{n-1}{2}}$; and when $n$ is even, $2^{n-1}-2^{\frac{n}{2}}$ is the maximum known nonlinearity, and it is conjectured that $NL(F)\leq 2^{n-1}-2^{\frac{n}{2}}$. It is well known that the inverse function has the known maximum nonlinearity. Particularly, its Walsh spectrum is characterized by the following lemma. \vspace{2mm}

{\bf Lemma 2.1.} \cite{Lachaud} For any positive integer $n$ and any $(a, b)\in F_{2^n}\times F_{2^n}^*$, the value of $|\sum_{x\in F_{2^n}}(-1)^{Tr(ax+bx^{-1})}|$ can be any integer divisible by 4 in the range $[-2^{n/2+1}+1, 2^{n/2+1}+1]$. \vspace{2mm}

Now we give the definition of the differential uniformity of a function $F$. \vspace{2mm}

\par {\bf Definition 2.2.} For an $(n,n)$-function $F$ and $(a, b)\in F_{2^n}^*\times F_{2^n}$, we denote by $\delta_F(a,b)$ the number of solutions of the equation $F(x+a)+F(x)=b$. The multiset $\{*~\delta_F(a,b)~|~(a, b)\in F_{2^n}^*\times F_{2^n}~*\}$ is called the differential spectrum of $F$, and the maximum value $\delta$ of this set is called the differential uniformity of $F$, or call $F$ differentially $\delta$-uniform. \vspace{2mm}

Particularly, $F$ is called almost perfect nonlinear (APN) if $\delta=2$. Note that the possible minimum value of $\delta$ equals 2 when $n$ is even, since both or neither $x$ and $x+a$ are solutions of $F(x+a)+F(x)=b$, for any $x\in F_{2^n}$.

Dillon proposed the switching method to obtain differentially low uniform functions from the known ones \cite{Dillon}. Some new APN functions were found following this method \cite{Budaghyan,Edel}. For a permutation $F$ of $F_{2^n}$ and $\nu\in F_{2^n}$, the functions with the form $F(x)+\nu f(x)$, where $f$ is a Boolean function (that is, an $(n, 1)$-function), are called switching neighbors of $F$ in the narrow sense. Recently, some new differentially 4-uniform permutations were found in the switching neighbors of the multiplicative inverse function in the narrow sense. For instance, the Qu-Tan-Tan-Li function \cite{Qu}, the compositional inverse of the Tang-Carlet-Tang function \cite{Tang} and the first class of the Zha-Hu-Sun function \cite{Zha} listed before are all of this type. In this paper, we shall present a new class of such differentially 4-uniform permutations.

Let $F, G$ be two $(n,n)$-functions. $F$ and $G$ are called to be extended affine (EA) equivalent if $F=L_2\circ G\circ L_1+A$ for some affine permutations $L_1, L_2$ over $F_{2^n}$ and some affine function $A$. $F$ and $G$ are called to be Carlet-Charpin-Zinoviev (CCZ) equivalent if there exists some affine automorphism $L=(L_1, L_2)$ of $F_{2^n}\times F_{2^n}$, where $L_1, L_2: F_{2^n}\times F_{2^n}\rightarrow F_{2^n}$ are affine functions, such that $y=G(x)$ if and only if $L_2(x, y)=F(L_1(x, y))$. It is well known that EA equivalence implies CCZ equivalence, but not for the converse. Moreover, CCZ equivalence and EA equivalence preserve the extended Walsh spectrum and the differential spectrum, and EA equivalence also preserves the algebraic degree when $n\geq 2$.

For a permutation $G: F_{2^n}\rightarrow F_{2^n}$, its compositional inverse function $F$ is defined by $G\circ F=F\circ G=id,$ where $id$ represents the identity mapping on $F_{2^n}$. And it has been proved that $G$ and $F$ are CCZ-equivalent.

Below we present some lemmas which are needed in the sequel. \vspace{2mm}

\par {\bf Lemma 2.3.}\label{1lem}\cite{MacWilliams} Let $n$ be a positive integer. For any $a,b,c\in F_{2^n}, ab\neq 0,$ the equation
\begin{eqnarray*} ax^2+bx+c=0
\end{eqnarray*}
has 2 solutions in $F_{2^n}$ if and only if $tr_1^n(ac/b^2)=0$. \vspace{2mm}

\par {\bf Lemma 2.4.}\label{2lem}\cite{Lachaud} Let $b\in F_{2^n}\setminus F_2$. Then $Tr\bigg(\frac{1}{b}\bigg)=0$ if and only if there exists some $\alpha\in F_{2^n}^*$ such that $b=\alpha+\alpha^{-1}$. \vspace{2mm}

\par {\bf Lemma 2.5.}\label{3lem}\cite{Qu} If $b=1+\alpha+\alpha^{-1}$ for some $\alpha\in F_{2^n}^*$, then the equation
\begin{eqnarray*} x^2+\frac{1}{b}x+\frac{1}{b(b+1)}=0
\end{eqnarray*} has two roots
\begin{eqnarray*} x_1=\frac{1}{1\!+\!\alpha\omega\!+\!(\alpha\omega)^{-1}}, ~x_2=\frac{1}{1\!+\!\alpha\omega^2\!+\!(\alpha\omega^2)^{-1}},
\end{eqnarray*} where $\omega\in F_{2^n}$ with order 3. \vspace{2mm}

\section{New differentially 4-uniform permutations}

Let $n\geq 6$ be an even integer and $V\subseteq F_{2^n}$ be a union of some pairs of elements $x$ and $\frac{x}{x+1}$ such that $Tr(x)=Tr\bigg(\frac{1}{x+1}\bigg)=1$ (here $V$ can be empty).

Let
\begin{eqnarray*}
W=\bigg\{x\in F_{2^n}~\bigg|~Tr(x)=Tr\bigg(\frac{1}{x+1}\bigg)=0\bigg\}
\end{eqnarray*}
and $U=V\cup W$. Define an $(n, n)$-function $G$ on $F_{2^n}$ as follows:
\begin{eqnarray*}
G(x)=\begin{cases}x^{-1}+1, x\in U;\\
x^{-1}, x\in F_{2^n}\setminus U.\end{cases}\end{eqnarray*}

Then the function $G$ can be written as
$G(x)=x^{-1}+\delta_U(x),$ where
\begin{eqnarray*}
\delta_U(x)=\begin{cases}1, x\in U;\\
0, x\in F_{2^n}\setminus U.\end{cases}\end{eqnarray*} \vspace{2mm}

{\bf Proposition 3.1.}
The function $G(x)$ is a permutation over $F_{2^n}$.

{\bf Proof.}  Let
\begin{eqnarray*}\varphi(x)=\begin{cases}\frac{1}{x^{-1}\!+\!1}, x\in U;\\
x, x\in F_{2^n}\setminus U.
\end{cases}\end{eqnarray*}

Note that we have
\begin{eqnarray*}\frac{1}{x^{-1}\!+\!1}=\begin{cases}1, x=0;\\
\frac{x}{x+1}, x\neq 0
\end{cases}\end{eqnarray*} and thus
\begin{eqnarray*}
&&Tr\bigg(\frac{1}{x^{-1}\!+\!1}\bigg)=Tr\bigg(\frac{1}{x\!+\!1}\bigg),\\
&&Tr\bigg(\frac{1}{\frac{1}{x^{-1}\!+\!1}\!+\!1}\bigg)=Tr(x\!+\!1)=Tr(x),
\end{eqnarray*} then it is easy to see that $\frac{1}{x^{-1}+1}\in U$ for any $x\in U$.
Hence $\varphi$ permutes $U$ and $F_{2^n}\setminus U$ respectively.
 Therefore, $\varphi(x)$ is a permutation over $F_{2^n}$, and thus
 $G(x)=\varphi(x)^{-1}$ is also a permutation over $F_{2^n}$. This completes the proof. \vspace{2mm}

Denote by $V_{M}$ the union of all possible sets $V$, which is the largest such set. Then it is clear that
\begin{eqnarray*} V_{M}=\bigg\{x\in F_{2^n}~\bigg|~Tr(x)=Tr\bigg(\frac{1}{x\!+\!1}\bigg)=1\bigg\}.
\end{eqnarray*}

{\bf Remark 3.2.}
Let $U=V_M\cup W$, then it is easy to see that $G_M(x)=x^{-1}+\delta_U(x)=x^{-1}+Tr(x+(x+1)^{-1})+1$, which is a translation of the 1st class function in \cite{Qu}. \vspace{2mm}

{\bf Remark 3.3.} Let $U=V$, then the function $G=x^{-1}+\delta_U(x)$ is the compositional inverse of some function constructed by \cite{Tang}. \vspace{2mm}

Now we consider the differentially uniformity of function $G$. For any $(a,b)\in F_{2^n}^*\times F_{2^n}$, the equation
\begin{eqnarray}\label{eq1}
G(x+a)+G(x)=b\end{eqnarray} is equivalent to the following two cases:

Case 1: Both or neither of $x$ and $x+a$ belong to $U$. In this case we get
\begin{eqnarray}\label{eq2}
x^{-1}+(x+a)^{-1}=b,\end{eqnarray}

Case 2: Exactly one of $x$ and $x+a$ belongs to $U$. In this case we get
\begin{eqnarray}\label{eq3}
x^{-1}+(x+a)^{-1}=b+1.\end{eqnarray}

Moreover, if $x\neq 0,a$, then Equation (\ref{eq2}) is equivalent to

\begin{eqnarray}\label{eq22}
bx^{2}+abx+a=0,
\end{eqnarray}

and Equation (\ref{eq3}) is equivalent to
\begin{eqnarray}\label{eq33}
(b+1)x^{2}+a(b+1)x+a=0.
\end{eqnarray}

To prove that the function $G(x)=x^{-1}+\delta_U(x)$ is differentially 4-uniform, we need the following results. \vspace{2mm}

\par {\bf Lemma 3.4.} Let $n$ be an even integer, then $Tr\bigg(\frac{1}{x+1}\bigg)=0$ if and only if $x=\frac{1}{1+\alpha+\alpha^{-1}}$ for some $\alpha\in F_{2^n}$, i.e.,
\begin{eqnarray*}
\bigg\{\frac{1}{1\!+\!\alpha\!+\!\alpha^{-1}}~\bigg|~\alpha\in F_{2^n}\bigg\}=\bigg\{x\in F_{2^n}~\bigg|~Tr\bigg(\frac{1}{x\!+\!1}\bigg)=0\bigg\}.
\end{eqnarray*}

{\bf Proof.} On the one hand, for any $\frac{1}{1+\alpha+\alpha^{-1}}$ in the left hand side set, one calculates
\begin{eqnarray*}
Tr\bigg(\frac{1}{\frac{1}{1\!+\!\alpha\!+\!\alpha^{-1}}\!+\!1}\bigg)&=&Tr\bigg(\frac{1\!+\!\alpha\!+\!\alpha^{-1}}{\alpha\!+\!\alpha^{-1}}\bigg)\\
&=&Tr(1)\!+\!Tr\bigg(\frac{1}{\alpha\!+\!\alpha^{-1}}\bigg)\\
&=&0,
\end{eqnarray*} where the last identity is due to $n$ being even and Lemma 2.4. So the left hand side set is contained in the right hand side set.

On the other hand, note that both sets must have the same number of elements, i.e. $2^{n-1}$, since the mapping $x+x^{-1}: F_{2^n}\rightarrow F_{2^n}$ is 2-to-1. Hence the two sets must be the same. This completes the proof. \vspace{2mm}

\par {\bf Proposition 3.5.} With the preceding notations as in Lemma 2.5, one has:

(1) $Tr\bigg(\frac{1}{x_1+1}\bigg)=Tr\bigg(\frac{1}{x_2+1}\bigg)=0$, and

(2) $Tr(x_1)+Tr(x_2)=Tr\bigg(\frac{1}{1+\alpha+\alpha^{-1}}\bigg)=Tr(b^{-1})$.

{\bf Proof.} (1) can be immediately obtained by Lemma 3.4.

For (2), one computes \begin{eqnarray*} &&Tr(x_1)\!+\!Tr(x_2)\\
=&&Tr\bigg(\frac{1}{1\!+\!\alpha\omega\!+\!(\alpha\omega)^{-1}}\!+\!\frac{1}{1\!+\!\alpha\omega^2\!+\!(\alpha\omega^2)^{-1}}\bigg)\\
=&&Tr\bigg(\frac{1}{1\!+\!\alpha\!+\!\alpha^{-1}}\bigg)\\
=&&Tr(b^{-1}),
\end{eqnarray*} where the second identity is based on
\begin{eqnarray*} \frac{1}{1\!+\!\alpha\omega\!+\!(\alpha\omega)^{-1}}\!+\!\frac{1}{1\!+\!\alpha\omega^2\!+\!(\alpha\omega^2)^{-1}}=\frac{1}{1\!+\!\alpha\!+\!\alpha^{-1}}.
\end{eqnarray*} This completes the proof. \vspace{2mm}

Now we give our main result. \vspace{2mm}

\par {\bf Theorem 3.6.} The permutation $G$ is differentially 4-uniform.

{\bf Proof.} Need to prove that Equation (\ref{eq1}) has at most 4 solutions.

Firstly, suppose $a\in U$. Then we have $Tr(a)=Tr\bigg(\frac{1}{a+1}\bigg)=1$ or $Tr(a)=Tr\bigg(\frac{1}{a+1}\bigg)=0$, according to $a\in V$ or $a\in W$ respectively.

When $ab=1$, then $0, a$ are two solutions of Case 1, and Case 1 has at most 4 solutions. In Case 2, Equation (\ref{eq3}) is equivalent to Equation (\ref{eq33}), we have
\begin{eqnarray*}x^2+ax+\frac{a^2}{a+1}=0,
\end{eqnarray*} or equivalently
\begin{eqnarray*}x^2+\frac{1}{b}x+\frac{1}{b(b+1)}=0.
\end{eqnarray*} If $a\in V$, then according to Lemma 2.3, Equation (\ref{eq33}) has no root, since $Tr\bigg(\frac{a^2}{a+1}\cdot\frac{1}{a^2}\bigg)=Tr\bigg(\frac{1}{a+1}\bigg)=1$.
Else if $a\in W$, then $Tr(a)=Tr\bigg(\frac{1}{a+1}\bigg)=0$, equivalently $Tr\bigg(\frac{1}{b}\bigg)=Tr\bigg(\frac{b}{b+1}\bigg)=Tr\bigg(\frac{1}{b+1}\bigg)=0$, and thus Equation (\ref{eq33}) has two roots. Moreover, by Lemma 2.4 there exists some $\alpha\in F_{2^n}^*$ such that
\begin{eqnarray*}
b=1+\alpha+\alpha^{-1}.
\end{eqnarray*}
According to Lemma 2.5, the two roots of Equation (\ref{eq33}) are exactly
\begin{eqnarray*} x_1=\frac{1}{1\!+\!\alpha\omega\!+\!(\alpha\omega)^{-1}}, ~x_2=x_1\!+\!a=\frac{1}{1\!+\!\alpha\omega^2\!+\!(\alpha\omega^2)^{-1}},
\end{eqnarray*}
where $\omega\in F_{2^n}$ with order 3. By Proposition 3.5 we have
\begin{eqnarray*}Tr\bigg(\frac{1}{x_1\!+\!1}\bigg)=Tr\bigg(\frac{1}{x_2\!+\!1}\bigg)=0,
\end{eqnarray*} and
\begin{eqnarray*} Tr(x_1)\!+\!Tr(x_2)=Tr(b^{-1})=Tr(a)=0.
\end{eqnarray*}  As a result, we get $Tr(x_1)=Tr(x_2)$, thus $x_1$ and $x_2$ are not solutions of Case 2. In fact, if $Tr(x_1)=Tr(x_2)=0$, then both $x_1$ and $x_2$ belong to $U$. While if $Tr(x_1)=Tr(x_2)=1$, then neither of $x_1$ and $x_2$ belongs to $U$.

When $ab\neq 1$, then Equations (\ref{eq2}) and (\ref{eq3}) are equivalent to Equations (\ref{eq22}) and (\ref{eq33}). Therefore, Equation (\ref{eq1}) has at most 4 solutions, since the sum of numbers of solutions of Equations (\ref{eq22}) and (\ref{eq33}) is at most 4.

Secondly, suppose $a\not\in U$.

When $a(b+1)=1$, then $0, a$ are two solutions of Case 2, and Case 2 has at most 4 solutions. In Case 1, Equation (\ref{eq2}) is equivalent to Equation (\ref{eq22}), we have
\begin{eqnarray*}x^2+ax+\frac{a^2}{a+1}=0,
\end{eqnarray*} or equivalently
\begin{eqnarray*}x^2+\frac{1}{b+1}x+\frac{1}{b(b+1)}=0.
\end{eqnarray*} If $Tr\bigg(\frac{1}{a+1}\bigg)=1$, then according to Lemma 2.3, Equation (\ref{eq22}) has no root, since $Tr\bigg(\frac{a^2}{a+1}\cdot\frac{1}{a^2}\bigg)=Tr\bigg(\frac{1}{a+1}\bigg)=1$.
Else if $Tr\bigg(\frac{1}{a+1}\bigg)=0$, equivalently $Tr\bigg(\frac{1}{b}\bigg)=0$, then Equation (\ref{eq22}) has two roots, and it holds $Tr(a)=1$, since $a\not\in W$.

Moreover, by Lemma 2.4 there exists some $\alpha\in F_{2^n}^*$ such that
\begin{eqnarray*}
b=\alpha+\alpha^{-1}.
\end{eqnarray*}
According to Lemma 2.5, the two roots of Equation (\ref{eq22}) are exactly
\begin{eqnarray*} x_1=\frac{1}{1\!+\!\alpha\omega\!+\!(\alpha\omega)^{-1}}, ~x_2=\frac{1}{1\!+\!\alpha\omega^2\!+\!(\alpha\omega^2)^{-1}}.
\end{eqnarray*}
Again from Proposition 3.5 we can get
\begin{eqnarray*} Tr\bigg(\frac{1}{x_1\!+\!1}\bigg)=Tr\bigg(\frac{1}{x_2\!+\!1}\bigg)=0,
\end{eqnarray*} and also
\begin{eqnarray*} Tr(x_1)+Tr(x_2)=Tr((b+1)^{-1})=Tr(a)=1.
\end{eqnarray*} As a result, one has $Tr(x_1)=Tr(x_2)+1$, which indicates that exactly one of $x_1$ and $x_2$ belongs to $U$, hence $x_1$ and $x_2$ are not solutions of Case 1. Therefore, Equation (\ref{eq1}) has at most 4 solutions.

When $a(b+1)\neq 1$, then Equations (\ref{eq2}) and (\ref{eq3}) are equivalent to Equations (\ref{eq22}) and (\ref{eq33}). Therefore, Equation (\ref{eq1}) has at most 4 solutions, since the sum of numbers of solutions of Equations (\ref{eq22}) and (\ref{eq33}) is at most 4. \vspace{2mm}

\section{Other Cryptographic Properties}

\subsection{Enumeration}

First we give the following estimation on $|V_{M}|$ and $|W|$, the cardinality of the sets $V_{M}$ and $W$. \vspace{2mm}

{\bf Proposition 4.1.} For even $n\geq 6$, we have $2^{n-2}-2^{n/2-1}\leq |V_{M}|=|W|\leq 2^{n-2}+2^{n/2-1}$.

{\bf Proof.} On the one hand, note that both $Tr(x)$ and $Tr((x+1)^{-1})$ are balanced Boolean functions, then we immediately get $|V_{M}|=|W|$, since for any two balanced Boolean functions $f, g: F_{2^n}\rightarrow F_2$, it holds ${\rm wt}(fg)={\rm wt}((f+1)(g+1))$, where ${\rm wt}(f)$ represents the Hamming weight of $f$.

On the other hand, we have
\begin{eqnarray*}V_{M}\cup W=\bigg\{x\in F_{2^n}~\bigg|~Tr(x)+Tr\bigg(\frac{1}{x+1}\bigg)=0\bigg\},
 \end{eqnarray*} and by Lemma 2.1, one gets that $|\sum_{x\in F_{2^n}}(-1)^{Tr(x+x^{-1})}|\leq 2^{n/2+1}$, or equivalently that $|\sum_{x\in F_{2^n}}(-1)^{Tr(x)+Tr((x+1)^{-1})}|\leq 2^{n/2+1}$ by changing $x$ to $x+1$. Therefore, we obtain
\begin{eqnarray*} 2^{n-1}-2^{n/2}\leq |V_{M}|+|W|\leq 2^{n-1}+2^{n/2}.
\end{eqnarray*} Combining with $|V_{M}|=|W|$, then the result follows.
This completes the proof. \vspace{2mm}

It is obvious then from Proposition 4.1 that in our construction the cardinality of $U$ satisfies $2^{n-2}-2^{n/2-1}\leq|U|\leq 2^{n-1}+2^{n/2}$. And since the size of $V$ can be any even integer ranging from 0 to $|V_{M}|$, so according to Proposition 4.1, there are at least $2^{2^{n-3}-2^{n/2-2}}$ different sets $V$, and thus there are at least $2^{2^{n-3}-2^{n/2-2}}$ different sets $U$. Moreover, in \cite{Tang} the authors computed the exact numbers of sets $V$ by computer for even integer $n$ ranging from 6 to 20. Therefore, we can also obtain the corresponding numbers of sets $U$ in our construction for even integer $n$ ranging from 6 to 20, and we list them in Table 1.

\begin{table}[h]\label{b0}
\caption{The exact numbers N of functions in our construction \cite{Tang}} \centering
\begin{tabular}{|c|c|c|c|c|c|c|c|c|}
\hline
$n$ & 6 & 8& 10 & 12& 14 & 16& 18 & 20\\
\hline
N & $2^7$ & $2^{36}$& $2^{121}$ & $2^{518}$& $2^{2059}$ & $2^{8136}$& $2^{32893}$ & $2^{130922}$\\
\hline
\end{tabular}
\end{table}

\subsection{Algebraic degree}

In \cite{Qu}, the authors characterized the algebraic degree of a large number of functions: \vspace{2mm}

{\bf Lemma 4.2.}\cite{Qu} Let $n$ be an integer, and $G(x)=x^{-1}+Tr(F(x))$ be a permutation over $F_{2^n}$. Then $G$ has optimal algebraic degree, i.e. $\deg(F)=n-1.$

According to Lemma 4.2, we immediately arrive at \vspace{2mm}

{\bf Corollary 4.3.} For every even integer $n\geq 6,$ all the functions $G$ in our construction have optimal algebraic degree $n-1.$ \vspace{2mm}

\subsection{Nonlinearity}

It has been proved that a special class of functions all have high nonlinearity: \vspace{2mm}

{\bf Theorem 4.4.}\cite{Qu} Let $n\geq 4$ be an integer, and $F(x)=x^{-1}+f(x)$, where $f: F_{2^n}\rightarrow F_{2}$ is a Boolean function such that $f\bigg(\frac{1}{x}\bigg)=f\bigg(\frac{1}{x+1}\bigg)$ holds for any $x\in F_{2^n}$. Then we have
\begin{eqnarray*}
NL(F)\geq 2^{n-2}-\frac{1}{4}\lfloor2^{\frac{n}{2}+1}\rfloor-1.
\end{eqnarray*}

We can use the above theorem to obtain the following corollary. \vspace{2mm}

{\bf Corollary 4.5.} For every even integer $n\geq 6,$ all the functions $G$ in our construction satisfy
\begin{eqnarray*}
NL(G)\geq 2^{n-2}-2^{\frac{n}{2}-1}-1.
\end{eqnarray*}

{\bf Proof.}
By definition, all the functions $G$ in our construction satisfy $G(x)=x^{-1}+\delta_U(x)$, where $\delta_U: F_{2^n}\rightarrow F_{2}$ is a Boolean function such that $\delta_U(x)=1$ if and only if $x\in U$. Besides, we have proved that the set $U$ satisfies $\frac{1}{x^{-1}+1}\in U$ for any $x\in U$. Therefore, it holds
$\delta_U(x)=\delta_U\bigg(\frac{1}{x^{-1}+1}\bigg)$, or equivalently $\delta_U\bigg(\frac{1}{x}\bigg)=\delta_U\bigg(\frac{1}{x+1}\bigg)$, for any $x\in F_{2^n}$, and thus the result follows by Theorem 4.4. \vspace{2mm}

We define two subsets of $V_{M}$. Let $V_{0}=\{x\in V_{M}~|~Tr\bigg(\frac{1}{x}\bigg)=0\}$, $V_{1}=\{x\in V_{M}~|~Tr\bigg(\frac{1}{x}\bigg)=1\}$. It is obvious that $V_{M}=V_{0}\cup V_{1}$. In what follows, we will give some improved lower bounds for the special cases $V=V_{0}, V_{1}$ and $\emptyset$ respectively.

We need the following results: \vspace{2mm}

{\bf Lemma 4.6.} \cite{Tang} For any positive integer $n$, any $b\in F_{2^n}$ and any Boolean function $f$ defined on $F_{2^n}$, it holds
\begin{eqnarray*}
|\sum_{x\in F_{2^n}, Tr(x)=c}(-1)^{f(x)+Tr(bx)}|\leq \max_{a\in F_{2^n}}|\sum_{x\in F_{2^n}}(-1)^{f(x)+Tr(ax)}|,
\end{eqnarray*} where $c\in F_2$. \vspace{2mm}

{\bf Lemma 4.7.} \cite{Qu} For any positive integer $n$ and any $(a, b)\in F_{2^n}\times F_{2^n}^*$, it holds
\begin{eqnarray*}
|\sum_{x\in F_{2^n}}(-1)^{Tr(ax+bx^{-1}+(x+1)^{-1})}|\leq 2\lfloor 2^{n/2+1}\rfloor\!+\!4.
\end{eqnarray*} \vspace{2mm}

{\bf Lemma 4.8.} For any even integer $n$ and any $(a, b)\in F_{2^n}^*\times F_{2^n}^*$, it holds
\begin{eqnarray*}
|\sum_{x\in F_{2^n}, Tr(x)=c}(-1)^{Tr(ax+bx^{-1}+(x+1)^{-1})+Tr(x^{-1})Tr((x+1)^{-1})}|\leq 6\cdot 2^{n/2}\!+\!4,
\end{eqnarray*}  where $c\in F_2$.

{\bf Proof.} For any fixed $(a, b)\in F_{2^n}^*\times F_{2^n}^*$, by Lemma 4.6, we have
\begin{eqnarray*}
&&|\sum_{x\in F_{2^n}, Tr(x)=c}(-1)^{Tr(ax+bx^{-1}+(x+1)^{-1})+Tr(x^{-1})Tr((x+1)^{-1})}|\\
&\leq& \max_{\widetilde{a}\in F_{2^n}}|\sum_{x\in F_{2^n}}(-1)^{Tr(\widetilde{a}x+bx^{-1}+(x+1)^{-1})+Tr(x^{-1})Tr((x+1)^{-1})}|,
\end{eqnarray*} while
\begin{eqnarray*}
&&|\sum_{x\in F_{2^n}}(-1)^{Tr(\widetilde{a}x+bx^{-1}+(x+1)^{-1})+Tr(x^{-1})Tr((x+1)^{-1})}|\\
&=&|\sum_{x\in F_{2^n}}(-1)^{Tr(\widetilde{a}x^{-1}+bx+(x+1)^{-1})+Tr(x)Tr((x+1)^{-1})}| \\
&=&|\sum_{x\in F_{2^n},Tr(x)=0}(-1)^{Tr(\widetilde{a}x^{-1}+bx+(x+1)^{-1})}\\
&&~~~~+\sum_{x\in F_{2^n},Tr(x)=1}(-1)^{Tr(\widetilde{a}x^{-1}+bx)}|\\
&\leq& (2^{\frac{n}{2}\!+\!2}\!+\!4)\!+\!2^{\frac{n}{2}\!+\!1}~({\rm by~ Lemmas~4.7~and~2.1~resp.})\\
&=&6\cdot2^{\frac{n}{2}}\!+\!4.
\end{eqnarray*} This completes the proof. \vspace{2mm}

Let $V=V_{0}$, then the function in our construction is
\begin{eqnarray*}G_1(x)=x^{-1}\!+\!1\!+\!Tr\bigg(x\!+\!\frac{1}{x\!+\!1}\bigg)+Tr(x)Tr\bigg(\frac{1}{x}\bigg)Tr\bigg(\frac{1}{x\!+\!1}\bigg).
 \end{eqnarray*} And we have \vspace{2mm}

 {\bf Theorem 4.9.} For every even integer $n\geq 6,$ it holds
\begin{eqnarray*}
NL(G_1)\geq 2^{n-1}-5\cdot2^{\frac{n}{2}}-4.
\end{eqnarray*}

{\bf Proof.} It suffices to show that
\begin{eqnarray*}
|G_1^{\mathcal{W}}(a,b)|=|\sum_{x\in F_{2^n}}(-1)^{Tr(ax+bG_1(x))}|\leq 10\cdot 2^{\frac{n}{2}}+8,
\end{eqnarray*} for any $(a, b)\in F_{2^n}\times F_{2^n}^*$. In fact, we have
\begin{eqnarray*}
&&|\sum_{x\in F_{2^n}}(-1)^{Tr(ax+bG_1(x))}|\\
&=&|\sum_{x\in F_{2^n}}(-1)^{Tr(ax\!+\!b(x^{-1}\!+\!1\!+\!Tr(x\!+\!(x+1)^{-1})\!+\!Tr(x)Tr(x^{-1})Tr((x\!+\!1)^{-1})))}|\\
&=&|\sum_{x\in F_{2^n}}(-1)^{Tr(ax\!+\!bx^{-1})\!+\!Tr(b)(1\!+\!Tr(x\!+\!(x\!+\!1)^{-1})\!+\!Tr(x)Tr(x^{-1})Tr((x\!+\!1)^{-1}))}|.
\end{eqnarray*}

If $Tr(b)=0,$ then $|F^{\mathcal{W}}(a,b)|=|\sum_{x\in F_{2^n}}(-1)^{Tr(ax+bx^{-1})}| \leq 2^{\frac{n}{2}+1}$, by Lemma 2.1.

Else if $Tr(b)=1,$ we get
\begin{eqnarray*}
&&|G_1^{\mathcal{W}}(a,b)|\\
&=&|\sum_{x\in F_{2^n}}(-1)^{Tr((a+1)x+bx^{-1}+(x+1)^{-1})+Tr(x)Tr(x^{-1})Tr((x+1)^{-1})}|\\
&=&|\sum_{x\in F_{2^n},Tr(x)=0}(-1)^{Tr((a+1)x+bx^{-1}+(x+1)^{-1})}+\\
&&\sum_{x\in F_{2^n},Tr(x)=1}(-1)^{Tr((a+1)x+bx^{-1}+(x+1)^{-1})+Tr(x^{-1})Tr((x+1)^{-1})}|\\
&\leq&(2^{n/2\!+\!2}\!+\!4)\!+\!(6\cdot2^{n/2}\!+\!4)~({\rm by~ Lemmas~4.7~and~4.8})\\
&=&10\cdot2^{\frac{n}{2}}\!+\!8.
\end{eqnarray*} This completes the proof. \vspace{2mm}

When $V=V_{1}$, the function in our construction is
\begin{eqnarray*}G_2(x)=x^{-1}\!+\!1\!+\!Tr\bigg(x\!+\!\frac{1}{x\!+\!1}\bigg)\!+\!Tr(x)(Tr\bigg(\frac{1}{x}\bigg)\!+\!1)Tr\bigg(\frac{1}{x\!+\!1}\bigg).
 \end{eqnarray*} And similar to the case $V=V_{0}$, we can get the following lower bound for $NL(G_2)$. \vspace{2mm}

{\bf Theorem 4.10.} For every even integer $n\geq 6,$ it holds
\begin{eqnarray*}
NL(G_2)\geq 2^{n-1}-5\cdot2^{\frac{n}{2}}-4.
\end{eqnarray*}  \vspace{2mm}

When $V=\emptyset$, the function in our construction is
\begin{eqnarray*}G_3(x)=x^{-1}\!+\!1\!+\!Tr\bigg(x\!+\!\frac{1}{x\!+\!1}\bigg)\!+\!Tr(x)Tr\bigg(\frac{1}{x\!+\!1}\bigg).
 \end{eqnarray*} And we have \vspace{2mm}

 {\bf Theorem 4.11.} For every even integer $n\geq 6,$ it holds
\begin{eqnarray*}
NL(G_3)\geq 2^{n-1}-3\cdot2^{\frac{n}{2}}-2.
\end{eqnarray*}

{\bf Proof.} It suffices to show that
\begin{eqnarray*}
|G_3^{\mathcal{W}}(a,b)|=|\sum_{x\in F_{2^n}}(-1)^{Tr(ax+bG_3(x))}|\leq 6\cdot 2^{\frac{n}{2}}+4,
\end{eqnarray*} for any $(a, b)\in F_{2^n}\times F_{2^n}^*$. In fact, we have
\begin{eqnarray*}
&&|\sum_{x\in F_{2^n}}(-1)^{Tr(ax+bG_3(x))}|\\
&=&|\sum_{x\in F_{2^n}}(-1)^{Tr(ax+b(x^{-1}+1+Tr(x+(x+1)^{-1})+Tr(x)Tr((x+1)^{-1})))}|\\
&=&|\sum_{x\in F_{2^n}}(-1)^{Tr(ax+bx^{-1})+Tr(b)(1+Tr(x+(x+1)^{-1})+Tr(x)Tr((x+1)^{-1}))}|.
\end{eqnarray*}

If $Tr(b)=0,$ then $|F^{\mathcal{W}}(a,b)|=|\sum_{x\in F_{2^n}}(-1)^{Tr(ax+bx^{-1})}|\leq 2^{\frac{n}{2}+1}$, by Lemma 2.1.

Else if $Tr(b)=1,$ we get
\begin{eqnarray*}
&&|G_3^{\mathcal{W}}(a,b)|\\
&=&|\sum_{x\in F_{2^n}}(-1)^{Tr((a+1)x+bx^{-1}+(x+1)^{-1})+Tr(x)Tr((x+1)^{-1})}|\\
&=&|\sum_{x\in F_{2^n},Tr(x)=0}(-1)^{Tr((a+1)x+bx^{-1}+(x+1)^{-1})}+\\
&&\sum_{x\in F_{2^n},Tr(x)=1}(-1)^{Tr((a+1)x+bx^{-1})}|\\
&\leq&(2^{n/2\!+\!2}\!+\!4)\!+\!2^{n/2\!+\!1}~({\rm by~ Lemmas~4.7~and~2.1~resp.})\\
&=&6\cdot2^{\frac{n}{2}}\!+\!4.
\end{eqnarray*} This completes the proof. \vspace{2mm}

\subsection{CCZ-inequivalence and computational results}

Generally, it is very hard to prove that two classes of functions are CCZ-inequivalent. People often turn to check whether their CCZ-invariant parameters, such as differential spectrum and extended Walsh spectrum, are the same. By a magma programme, we computed the nonlinearity and the differential spectrum of the newly discovered functions for $n=6$, and found that there are at least 13 CCZ-inequivalent classes, which are listed in Table 2 (excluding the 1st class function $G_M$ of \cite{Qu}), where we let $\xi$ be the default primitive element of $F_{2^6}$ in Magma version 2.19-10, $NL$ be the nonlinearity of a function, and $[a, b, c]$ be the differential spectrum of a function, which means that $0$ appears $a$ times, $2$ appears $b$ times and $4$ appears $c$ times respectively. Comparing with Table 3 of \cite{Tang}, one can find that among these 13 classes of functions in Table 2, the nonlinearity and differential spectrum of 6 classes of them (with *) are different from those in \cite{Tang}, which indicates that they are CCZ-inequivalent to the functions constructed by \cite{Tang}.

\begin{table}[h]\label{b2}
\caption{CCZ-inequivalent differentially 4-uniform permutations for $n=6$} \centering
\medskip
\begin{tabular}{|c|c|c|c|}
\hline
$V$ & $NL$ & Differential spectrum&\\
\hline
$\emptyset$ & $20$ & $[2235,1578,219]$&\\
\hline
$\{\xi^i~|~i\in\{21, 42\}\}$ & $20$ & $[2247,1554,231]$&\\
\hline
$\{\xi^i~|~i\in\{3, 53\}\}$ & $18$ & $[2253,1542,237]$&\\
\hline
$\{\xi^i~|~i\in\{3, 6, 43, 53\}\}$ & $18$ & $[2265,1518,249]$&\\
\hline
$\{\xi^i~|~i\in\{3, 12, 23, 53\}\}$ & $18$ & $[2259,1530,243]$&\\
\hline
$\{\xi^i~|~i\in\{3, 21, 42, 53\}\}$ & $18$ & $[2271,1506,255]$&\\
\hline
$\{\xi^i~|~i\in\{3, 12, 23, 29, 48, 53\}\}$ & $20$ & $[2253,1542,237]$&\\
\hline
$\{\xi^i~|~i\in\{3, 21, 24, 42, 46, 53\}\}$ & $20$ & $[2277,1494,261]$&*\\
\hline
$\{\xi^i~|~i\in\{3, 21, 29, 42, 48, 53\}\}$ & $18$ & $[2283,1482,267]$&*\\
\hline
$\{\xi^i~|~i\in\{3, 21, 33, 42, 53, 58\}\}$ & $18$ & $[2289,1470,273]$&*\\
\hline
$\{\xi^i~|~i\in\{3, 12, 21, 23, 29, 42, 48, 53\}\}$ & $20$ & $[2283,1482,267]$&*\\
\hline
$\{\xi^i~|~i\in\{3, 21, 29, 33, 42, 48, 53, 58\}\}$ & $18$ & $[2295,1458,279]$&*\\
\hline
$\{\xi^i~|~i\in\{3, 6, 21, 29, 42, 43, 48, 53\}\}$ & $20$ & $[2295,1458,279]$&*\\
\hline

\end{tabular}
\end{table}

In \cite{Tang}, three special classes of differentially 4-uniform permutations $F_1, F_2$ and $F_3$ were identified and proved to be CCZ-inequivalent to known differentially 4-uniform power functions and to quadratic functions, where
\begin{eqnarray*}
&&F_1(x)=\bigg(x+Tr\bigg(\frac{1}{x}\bigg)Tr\bigg(\frac{1}{x+1}\bigg)\bigg)^{-1},\\
&&F_2(x)=\bigg(x+(1+Tr(x))Tr\bigg(\frac{1}{x}\bigg)Tr\bigg(\frac{1}{x+1}\bigg)\bigg)^{-1},\\
&&F_3(x)=\bigg(x+Tr(x)Tr\bigg(\frac{1}{x}\bigg)Tr\bigg(\frac{1}{x+1}\bigg)\bigg)^{-1}.
\end{eqnarray*}

For our three special subclasses $G_1, G_2$ and $G_3$, we have also computed their differential spectrum for small $n=6, 8, 10$ and compared with that of $G_M$ constructed by \cite{Qu} (see Remark 3.2), and functions $F_1, F_2, F_3$. Although they are constructed by permuting the multiplicative inverse functions, the differential spectrum is different. The computational results are listed in Table 3. We also list the lower bounds on nonlinearity of these functions in Table 4, where the last row is the known maximum values of the nonlinearity.

\begin{table}[h]\label{b3}
\caption{The differentially spectrum of $G_M$, $G_i$ and $F_i$ for $n=6, 8, 10$} \centering
\medskip
\begin{tabular}{|cccc|}
\hline
 & $n=6$ & $n=8$ & $n=10$\\
 \hline
\!$G_1$\!&\!$[2253,1542,237]$\!& \!$[36735,24450,4095]$\!& \!$[586269,398790,62493]$\! \\
\hline
\!$G_2$\!&\!$[2247,1554,231]$\!& \!$[36495,24930,3855]$\!& \!$[589089,393150,65313]$\! \\
\hline
\!$G_3$\!&\!$[2235,1578,219]$\!& \!$[36159,25602,3519]$\!& \!$[581289,408750,57513]$\! \\
\hline
\!$G_M$\!&\!$[2301,1446,285]$\!& \!$[36735,24450,4095]$\!& \!$[589389,392550,65613]$\! \\
\hline
$F_1$& $[2289,1470,273]$&$[36207,25506,3567]$& $[580779,409770,57003]$ \\
\hline
$F_2$ &$[2277,1494,261]$& $[35199,27522,2559]$& $[566259,438810,42483]$\\
\hline
$F_3$ &$[2127,1794,111]$& $[36159,25602,3519]$& $[565839,439650,42063]$ \\
\hline
\end{tabular}
\end{table}

\begin{table}[h]\label{b2}
\caption{Lower bounds for nonlinearity and exact values for small $n$ (the last row is the known maximum nonlinearity) } \centering
\medskip
\begin{tabular}{|c|c|c|c|c|c|c|}
\hline
Functions & Lower bounds & $n=6$ & $n=8$ & $n=10$& $n=12$\\\hline
       $G_1$ &  $2^{n-1}-5\cdot2^{n/2}-4$ &20 &96 &450&1888\\
       \hline
       $G_2$ &  $2^{n-1}-5\cdot2^{n/2}-4$ &20 &100 &442&1910\\
       \hline
       $G_3$ &  $2^{n-1}-3\cdot2^{n/2}-2$ &20 &100 &446&1912\\
       \hline
       $G_M$ &  $2^{n-1}-2^{n/2+1}-2$ &20 &102 &454 &1928\\
       \hline
       $F_1$  &  $2^{n-1}-3\cdot2^{n/2}-2$ &20 &100 &442 &1902\\
       \hline
       $F_2$ &  $2^{n-1}-2^{n/2+2}-2$ &22 &102 &450 &1924\\
       \hline
       $F_3$  &  $2^{n-1}-2^{n/2+2}-2$ &22 &102 &444 &1892\\
\hline
 MAX &  $2^{n-1}-2^{n/2}$ &24 &112 &480 &1984\\
\hline

\end{tabular}
\end{table}

\vspace{2mm}

\section{Conclusion}

In this paper we presented a new construction of differentially 4-uniform permutations for any even $n\geq 6$, which gives many new CCZ-inequivalent functions. We also deduced a lower bound on the nonlinearity of our functions, and three improved lower bounds for three infinite subclasses.




%

\end{document}